# Spectroscopic study of native defects in the semiconductor to metal phase transition in $V_2O_5$ nanostructure

Raktima Basu[a)] and Sandip Dhara[a)]

*Surface and Nanoscience Division, Indira Gandhi Centre for Atomic Research, Homi Bhabha National Institute, Tamil Nadu 603102, India*

*Abstract*

Vanadium is a transition metal with multiple oxidation states and $V_2O_5$ is the most stable form among them. Besides catalysis, chemical sensing and photo-chromatic applications, $V_2O_5$ is also reported to exhibit a semiconductor to metal transition (SMT) at a temperature range of 530-560K. Even though, there are debates in using the term 'SMT' for $V_2O_5$, the metallic behavior above transition temperature and its origin are of great interests in the scientific community. In this study, $V_2O_5$ nanostructures were deposited on $SiO_2$/Si substrate by vapour transport method using Au as catalyst. Temperature dependent electrical measurement confirms the SMT in $V_2O_5$ without any structural change. Temperature dependent photoluminescence analysis proves the appearance of oxygen vacancy related peaks due to reduction of $V_2O_5$ above the transition temperature, as also inferred from temperature dependent Raman spectroscopic studies. The newly evolved defect levels in the $V_2O_5$ electronic structure with increasing temperature is also understood from the downward shift of the bottom most split-off conduction bands due to breakdown of *pdπ* bonds leading to metallic behavior in $V_2O_5$ above the transition temperature.

[a)]Authors to whom correspondence should be addressed. Electronic addresses: raktimabasu14@gmail.com, dhara@igcar.gov.in.

## I. INTRODUCTION

Vanadium oxides find tremendous attention because of their excellent structural flexibility and outstanding physical properties. $V_2O_5$, the most stable one among them with maximum oxidization state of +5, exhibits interesting structural, optical, and electrical properties and therefore is applicable in various fields, such as gas sensing,[1] optoelectronic switches,[2] rechargeable lithium batteries,[3] and photo-chromic devices.[4] Most of the vanadium oxides, namely $VO_2$, $V_2O_3$, and $V_6O_{13}$ undergo a semiconductor to metal transition (SMT) as a function of temperature.[5-7] $V_2O_5$ in its various forms of nanorods, thin films are also reported to exhibit a SMT around the transition temperature of 530K.[8,9] In a separate report, single crystal (001) surface facet of $V_2O_5$ is reported with a SMT at comparatively low temperature of 350 to 400K in the localized scanning tunneling microscopic measurement at ultrahigh vacuum.[10] However, there are arguments on the origin of the metallic behavior above the transition temperature. In different reports, the reduction of $V_2O_5$ to other lower ordered stoichiometric and non-stoichiometric oxides,[11] as well as a structural change from $\alpha$-$V_2O_5$ to $\gamma'$-$V_2O_5$,[12] are proposed to be responsible for the SMT.

In this present study, we report metallic behavior in $V_2O_5$ nanostructures above transition temperature of 530K without any structural change, by means of temperature dependent electrical measurements. The origin of the metallic behavior is discussed considering the observed emission peaks above transition temperature in the temperature dependent photoluminescence (PL) spectroscopic studies. The newly evolved PL emission above the transition temperature is understood in terms of creation of defect states leading to the modification of conduction bands in the $V_2O_5$ electronic structure invoking metallicity in the system.

## II. EXPERIMENTAL DETAILS

### A. Synthesis

$V_2O_5$ nanostructures were synthesized by vapor transport process using bulk $V_2O_5$ powder (Sigma-Aldrich, 99%) as source and Au as catalyst. Au thin film (2 nm) coated $SiO_2$/Si (100) was used as substrate and Ar was used as carrier gas. Au film was deposited on $SiO_2$/Si (100) film using thermal evaporation technique (12A4D, HINDHIVAC, India) under high vacuum ($10^{-6}$ mbar). The bulk $V_2O_5$ powder was placed in a high pure (99.99%) alumina boat at the center of the quartz tube reaction chamber, kept in a thermoelectrically controlled furnace. The substrate was kept 5 cm away from the source and perpendicular to the stream of Ar. The reaction chamber was evacuated up to $10^{-3}$ mbar with the help of a rotary pump. The temperature of the quartz tube was programmed to rise up to 500 °C and stabilize at that temperature for 10 min to form Au islands on the substrate and then increased to the optimized growth temperature (1173K) with 15 ° C min$^{-1}$ ramp rate. The synthesis was carried out at 1173K flowing 20 sccm of commercial Ar as carrier gas for 60 min.

**B. Characterizations**

The morphological analysis of the pristine sample was performed using a field emission scanning electron microscope (FESEM, SUPRA 55 Zeiss). The crystallographic studies were carried out with the help of glancing incidence X-ray diffractometer (GIXRD; Bruker D8) using a Cu Kα radiation source (λ = 1.5406 Å) with a glancing angle (θ) of 0.5°. Electrical properties were investigated in a voltage range of 2 V using two Au coated contact tips with the help of a source measurement unit (Agilent B2911A). A micro-Raman spectrometer (inVia, Renishaw, UK) was used with Ar$^+$ Laser (514.5 nm) as excitation source and a diffraction grating of 1800 gr.mm$^{-1}$ as monochromator to study the vibrational modes of the synthesized sample. A thermoelectrically cooled CCD camera was used as the detector for the Raman spectra in the back scattering configuration. The spectra were collected using a long working distance 50x objective with numerical aperture of 0.45. Absorption spectra in reflection geometry were recorded using an UV–Visible (UV-Vis) absorption spectrometer (Avantes) in the range of 300– 800 nm. A bare $SiO_2$/Si (100) wafer was used for reference to nullify the reflected contribution from the substrate. PL spectra were recorded with an excitation of 325 nm UV Laser and 2400 gr.mm$^{-1}$ grating with help of the same Raman spectrometer to understand the change in electronic states. In order to perform temperature dependent electrical and spectroscopic measurements, the samples were kept in a Linkam (THMS600) stage with an auto-controlled thermoelectric heating and cooling function within a temperature range of 80 to 650K.

**III. RESULTS AND DISCUSSION**

The typical FESEM image of as-grown sample shows crystallites of different sizes (Fig. 1(a)). The region with the early nucleation of oxide phase with Au NPs is also recorded in the typical FESEM image (Fig. 1(b)). The inset in figure 1(a) shows high magnification image of a typical single nanorod with an average diameter of 200-300 nm along with Au NPs. The typical diameter of the Au particles is found out ~ 40 nm (Fig. 1(b)). The GIXRD pattern (Fig. 1(c)) confirms the presence of pure $V_2O_5$ phase (ICCD 00−041−1426) with textured (001) plane along with (111) planes corresponding to Au (ICCD 00−004−0784).

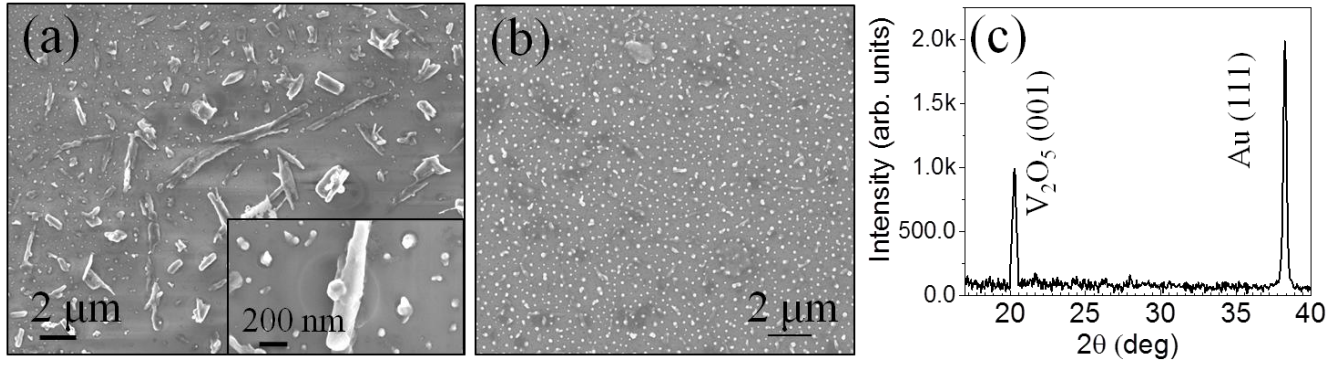

**FIG. 1.** (a) FESEM image of as grown nanostructures. Inset shows a typical single $V_2O_5$ nanorod of diameter ~ 200-300 nm and spherical Au NPs (b) FESEM image of the Au NPs in the early nucleation of oxide phase (c) GIXRD pattern of the as grown sample indicating crystallographic planes.

The space group of $V_2O_5$ is $P_{mmn}$ ($D_{2h}^{13}$).[13] Among twenty one group theoretically predicted Raman active modes for $V_2O_5$ at $\Gamma$ point, ($7A_g+7B_{2g}+3B_{1g}+4B_{3g}$),[14,15] we observed eleven Raman modes for the as-grown nanostructures (Fig. 2).

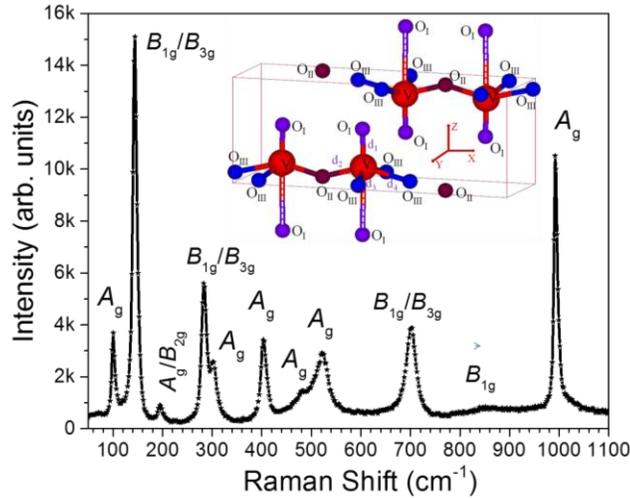

**FIG. 2.** Raman spectrum of as grown nanostructures. Inset shows a schematic diagram of the $V_2O_5$ unit cell. Van der Waals bond between V and $O_I$ is indicated by white dashed line.

The Raman peaks at 102 ($A_g$), 144 (either $B_{1g}$ or $B_{3g}$; $B_{1g}/B_{3g}$), 195 ($A_g/B_{2g}$), 283 ($B_{1g}/B_{3g}$), 301 ($A_g$), 403 ($A_g$), 483 ($A_g$), 523 ($A_g$), 701 ($B_{1g}/B_{3g}$), 850 ($B_{1g}$) and 993 ($A_g$) cm$^{-1}$ confirm the formation of pure $V_2O_5$ phase.[16] Orthorhombic $V_2O_5$ is composed by distorted $VO_5$ pyramids, sharing edges and corners. A schematic diagram of $V_2O_5$ unit cell is shown in the inset of figure 2. There are three structurally different oxygen atoms in each unit cell (denoted as $O_I$, $O_{II}$, and $O_{III}$ in the inset of Fig. 2). $O_I$ is the terminal (vanadyl) oxygen with two different bond lengths; strong and short V-O$_I$ bond is with length 1.577 Å ($d_1$) and large and weak one is with a bond length of 2.793 Å. The later bond is of Van der Waals type, which connects the two adjacent layers in the V$_2$O$_5$ structure. The two fold coordinated bridging oxygen (O$_{II}$) connects two adjacent V atoms with V-O$_{II}$ bond length of 1.78 Å ($d_2$). The ladder shaped O$_{III}$ atoms are the three-fold co-ordinated oxygen with three different V-O$_{III}$ bond lengths of 1.88 ($d_3$), 1.88 ($d_3$), and 2.02 Å ($d_4$).[17] V$_2$O$_5$ consists of layers connecting by a weak Van der Waals bond between V and O$_I$ atom of the adjacent layer (shown by white dashed line in figure). The highest frequency Raman mode at 993 cm$^{-1}$ arises due to the vibration of terminal oxygen atoms along Z direction and is a signature peak for V$_2$O$_5$.[18,19] The peak at 850 cm$^{-1}$ is predicted to originate because of antiphase stretching mode of V-O$_{II}$ bonds.[16] Displacement of O$_{III}$ atoms in *Y* and *X* directions generates Raman modes at 701 cm$^{-1}$ and 523 cm$^{-1}$, respectively. The V-O$_{II}$-V bending deformation along *Z* direction gives rise to Raman mode at 483 cm$^{-1}$. Modes at 403 and 283 cm$^{-1}$ can be attributed to oscillation of O$_I$ atoms along *X* and *Y* axes, respectively. One the other hand, displacement of O$_{II}$ atoms along *Z* axis gives rise to Raman peak at 301 cm$^{-1}$. The low frequency modes at 195, 144 and 102 cm$^{-1}$ correspond to the *X*, *Y* and *Z* displacements of the whole chain of V-O$_{III}$ bonds. The high intensity of 144 cm$^{-1}$ peak indicates the long range order of V-O layers in the *XY* plane.[19,20] The Raman mode at 850 cm$^{-1}$ is reported not to be observed experimentally due to the pseudo-centrosymmetric nature of V−OII−V bond. However, we could observe the mode because of possible surface enhanced Raman scattering due to the plasmonic effect of the Au NPs present (Fig. 1) in the sample.

The electrical measurement shows (Fig. 3(a)) that resistance decreases exponentially up to 450K with increasing temperature. Above 550K, resistance falls rapidly disobeying the semiconducting nature and the plot shows metallic behavior of increasing resistance with increase in the temperature above 650K. The proposed structural phase transition to the metastable γ′-V$_2$O$_5$ can be ruled out as γ′-V$_2$O$_5$ is reported to transform to semiconducting α-V$_2$O$_5$ phase above 613K. In the present study, however the temperature dependent electrical measurement shows (Fig. 3(a)) that the metallic character of the grown nanorods sustains above 650K, which contradicts the formation of γ′-V$_2$O$_5$. Raman spectroscopy is a nondestructive technique for structural as well as phase confirmation. So, temperature dependent Raman study was performed to address the issue in SMT for V$_2$O$_5$. Figure 3(b) shows Raman spectra taken at different temperatures ranging from 300 to 650K.

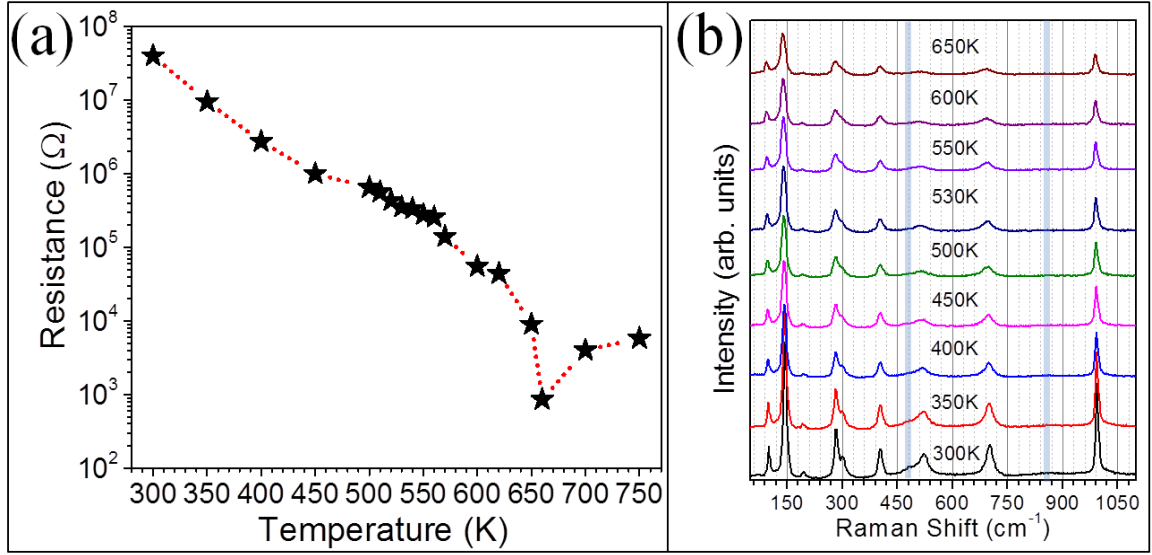

**FIG. 3.** (a) Change in the resistance with temperature in an ensemble of $V_2O_5$ nanostructure. Dotted line is guide to eye. (b) Raman spectra at different temperatures ranging from 300 to 650K. Shaded lines indicate the Raman modes disappearing above the transition temperature.

We observed Raman modes at 483 cm$^{-1}$ and 850 cm$^{-1}$ to disappear completely above temperature 523K (shaded lines in Fig. 3(b)). Disappearance of the Raman modes agrees well with the relaxed structure of orthorhombic $V_2O_5$ after reduction of oxygen.[8,21] The abrupt decrease in intensity of the highest frequency peak at 993 cm$^{-1}$ with increasing temperature also indicates loss of oxygen from the system. Structural transition from α-$V_2O_5$ to γ´-$V_2O_5$ above transition temperature[12] can be further eliminated, as the observed Raman modes above the transition temperature do not resemble with that of γ´-$V_2O_5$ amid the absence of characteristic Raman mode at 602 cm$^{-1}$ in the studied temperature range of 300-650K.[22]

$V_2O_5$ is reported as a semiconductor with a large energy gap of 3.3 eV; within this gap, however there exists two localized bands.[23] The schematic diagram of the electronic band structure of $V_2O_5$ is shown in figure 4(a). $V_2O_5$ is a semiconductor with an indirect band gap value of 2.1 eV corresponding to a transition from R to Γ point in the first Brillouin zone.[24] The direct band gap at Γ point is reported ~2.5 eV.[23,25] Two split-off bands with narrow bandwidth below the conduction band at Γ point are also reported due to the overlapped O 2p and V 3d bands (pdπ bonds).[26,27] We have carried out the UV-Vis absorption spectroscopic studies in the reflection geometry to understand the optical properties in the light of electronic transitions. Figure 4(b) shows UV–Vis absorption spectra of the sample. As there are Au NPs present in the sample, to investigate the role of Au in the absorption spectrum of $V_2O_5$, we performed the absorption studies for Au NPs as well as bulk $V_2O_5$ also. Au NPs of diameter ~ 40 nm show a strong absorption peak at 528 nm, whereas bulk $V_2O_5$ shows four peaks around 560 to 630 nm. A faint absorption edge at ~ 342 nm may indicate presence of theoretically forbidden interband transition in the bulk $V_2O_5$ powder.[23] The feature may have appeared because of the finite size of $V_2O_5$ powder and the measurements being performed at room temperature. On the other hand, $V_2O_5$ nanostructures in presence of Au NPs show two strong absorption peaks at 342 and 565 nm. The peak at 565 nm (2.19 eV) matches with the previous report and can be attributed as the indirect transition from R to Γ point.[23-27] The peak at 342 nm (3.63 eV) corresponding to the forbidden interband transition[28] of $V_2O_5$, may have shown its strong presence due to the plasmonic effect of Au NPs influencing the effective polarizability of the hybrid system.[29] Au NPs with diameter of 40 nm on $SiO_2$ matrix is reported to show luminescence ~3.4 eV corresponding to energy gap at L –symmetry point.[30]

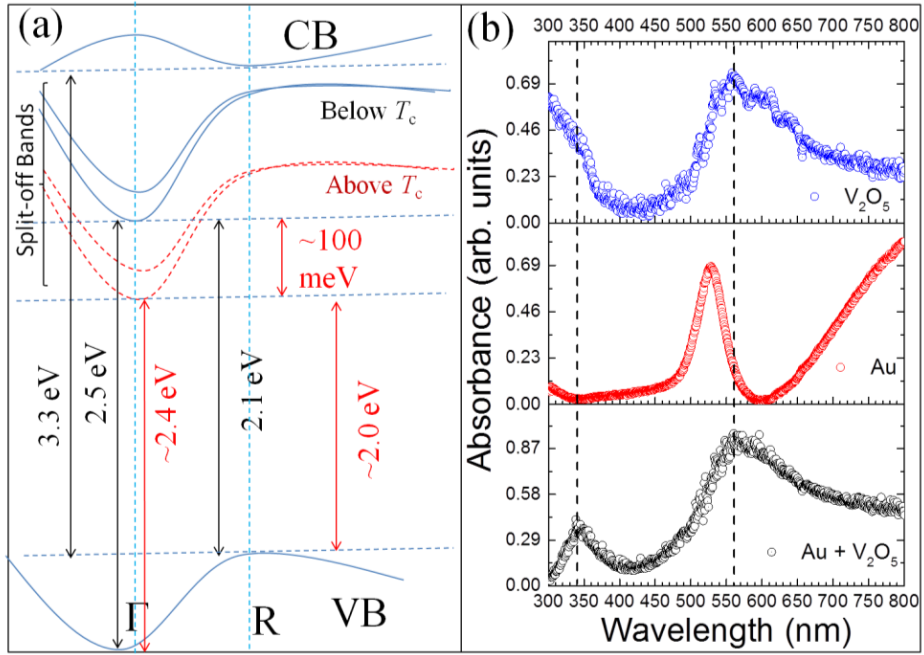

**FIG. 4.** (a) Schematic electronic band structure for $V_2O_5$ indicating the possible transition. Solid and dashed split-off curves denote the electronic states below and above the transition temperature ($T_c$), respectively. (b) UV–Vis absorption spectra of the grown sample, Au NPs and bulk $V_2O_5$.

For, further analysis we carried out temperature dependent PL studies to examine the changes in electronic states with temperature. Figure 5 shows PL spectra of $V_2O_5$ nanostructures with increasing temperature from 80 to 650K. We observed three sharp peaks at 2.1, 2.5 and 2.7 eV, recorded at 80K, which can be explained as an indirect transition from R to Γ point, a direct transition at Γ point and a direct transition at R point, respectively.[23,25] The observation of PL peaks for indirect transition in $V_2O_5$ is facilitated using the near resonance condition where laser excitation of 325 nm is selected close to the plasmonic assisted absorption peak of the hybrid system at 342 nm (Fig. 4(b)). At room temperature (300K), the peaks are blue sifted slightly ~ 30 meV. The trend continues upto 500K, which is quite expected with increasing temperature. Surprisingly above the transition temperature at 550K, three new peaks arise around 1.89, 2.0 and 2.4 eV along with the previously observed peaks. With increasing temperature, the split-off conduction bands in the $V_2O_5$ electronic structure are reported to approach deeper down from the conduction band around 70 to 100 meV at Γ point due to the breakdown of *pd*π bond between O and nearest V atoms (Fig. 4(a)).[8] The origin of the newly evolved peaks can be understood because of the indirect transition from R to Γ point (1.89 and 2.0 eV) and a direct transition at Γ point (2.4 eV) of the new electronic levels, caused by oxygen vacancies (considering dotted split-off curves in Fig. 4(a)).[8,31] With further increase in temperature upto 650K, the intensity of newly evolved low energy vacancy related peaks increases whereas, the intensity of band edge peaks decreases. Thus, the above observation supports the loss of oxygen from the system with increase in temperature,[32] as also observed in the temperature dependent Raman spectroscopic studies (Fig. 2). If one oxygen atom is removed from the structure, it leaves two electrons in the system, which in turn may give rise to the conductivity of the material. The breakdown of *pd*π bonds between O 2*p* and V 3*d* drives the electrons toward partially filled V 3*d* bands, which in turn increases the number of carriers in conduction band leading to the observed metallic behavior.

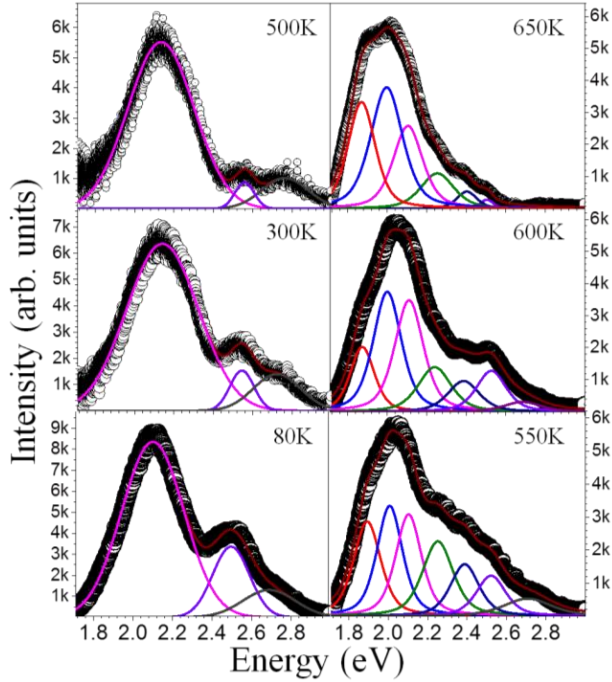

**FIG. 5.** PL spectra at different temperatures ranging from 80 to 650K. Peaks are fitted with Gaussian function. Symbols indicate data points, and lines indicate fitted curves.

## IV. CONCLUSION:

In conclusion, $V_2O_5$ nanostructures were grown by vapour transport mechanism using Au as catalyst. The Raman spectroscopic analysis and GIXRD measurements confirm the orthorhombic phase of the as grown material. Temperature dependent electrical measurements prove the SMT in $V_2O_5$ without any structural change. In temperature dependent Raman spectroscopic studies, the disappearance of Raman modes corresponding to specific V-O bonds above the transition temperature confirms the relaxation of $V_2O_5$ structure due to loss of oxygen without any global structural change. Appearance of new peaks in the PL spectra above transition temperature of 530K is understood in terms of the reduction of $V_2O_5$ due to oxygen vacancy. The formation of new defect states in $V_2O_5$ electronic structure above the transition temperature reveals the cause of metallic behavior arising from downward shift of the bottom most split-off conduction bands due to breakdown of *pdπ* bonds between O and nearest V atoms.

**Acknowledgements**

R. B. acknowledges the Department of Atomic Energy for allowing her to continue the research work. We thank R. Pandian and D. N. Sunitha of SND, IGCAR for their help in FESEM and GIXRD study.